\documentclass[letter,10pt,conference]{IEEEtran}

\usepackage{caption}

\DeclareCaptionLabelFormat{lc}{{#1}~#2}
\def\tablename{Table}

\ifCLASSINFOpdf
\else
\fi

 \usepackage{epsfig,endnotes}
\let\labelindent\relax
\usepackage{floatpag,enumitem}
\usepackage{hyphenat}

\usepackage{cite}
\usepackage{relsize}
 \usepackage[top=.75in, bottom=.85in, left=0.45in, right=0.45in]{geometry}
 \usepackage{tikz}

\newcommand{\de}{\stackrel{\text{def}}{=}}
\newcommand{\con}{\boldsymbol{\mid}}

\usepackage{multirow}
\usepackage{setspace}
\usepackage{mathrsfs}
\usepackage{bbm}
\usepackage{pifont}
\usepackage{amsfonts}

\usepackage{amsmath, amsthm}

\newenvironment{rcases}
  {\left.\begin{aligned}}
  {\end{aligned}\right\rbrace}
\usepackage{tabularx}
\usepackage{listings}
\usepackage{graphicx}
\usepackage{amssymb}
\usepackage{array}
\usepackage{fancyhdr}
\usepackage{cases}

 \usepackage{supertabular}
\usepackage{url}
\usepackage{hyperref}
\usepackage{fancyhdr}
\usepackage{mathrsfs}

\makeatletter
\def\@copyrightspace{\relax}
\makeatother

\usepackage{psfrag}
\usepackage{bbding}

\usepackage{float}

\usepackage{amsbsy}

\makeatletter
\providecommand{\leftsquigarrow}{%
  \mathrel{\mathpalette\reflect@squig\relax}%
}
\newcommand{\reflect@squig}[2]{%
  \reflectbox{$\m@th#1\rightsquigarrow$}%
}
\makeatother

\usepackage{algorithm}
 \usepackage{algorithmic}

\makeatletter
\newcommand{\newalgname}[1]{%
  \renewcommand{\ALG@name}{#1}%
}

\newcommand {\C} {{\rm I\kern-5.5pt C}}

\newcommand{\bp}[1]{{\mathbb{P}}\left[{#1}\right]}
\newcommand{\bigp}[1]{{\mathbb{P}}\big[{#1}\big]}

\newcommand{\fsquare}{\vrule height6pt width7pt depth1pt}   
\newcommand{\pfe}{\hfill\fsquare}             

\def\centerhack#1{\hbox to 0pt{\hss\footnotesize #1\hss}}
\def\centerhackn#1{\hbox to 0pt{\hss #1\hss}}
\def\dchack#1{\vbox to 0pt{\vss{\hbox to 0pt{\hss#1\hss}}\vss}}
\newcommand{\pr}{\mathbb{P}} 

\usepackage{etoolbox}
\AtBeginEnvironment{align}{\setcounter{subeqn}{0}}
\newcounter{subeqn} %

\usetikzlibrary{positioning}

\newcounter{mysub}

\newtheorem{defn}{Definition}

\newtheorem{lem}{Lemma}
\newtheorem{thm}{Theorem}

\newtheorem{cor}{Corollary}

\newtheorem*{proposition1.1}{Proposition 1.1}
\newtheorem*{proposition1.2}{Proposition 1.2}
\newtheorem*{proposition1.3}{Proposition 1.3}
\newtheorem*{proposition2.1}{Proposition 2.1}
\newtheorem*{proposition2.2}{Proposition 2.2}
\usepackage{subfigure}

\IEEEoverridecommandlockouts

\begin{document}

\title{Relations Among Different  Privacy Notions}

%

\author{\IEEEauthorblockN{Jun Zhao}
\IEEEauthorblockA{\href{junzhao@alumni.cmu.edu}{\texttt{junzhao@alumni.cmu.edu}}\thanks{The author Jun Zhao obtained his PhD from Carnegie Mellon University, Pittsburgh, PA 15213, USA, where he was with the Cybersecurity Lab (CyLab). He was a postdoctoral scholar with Arizona State University, Tempe, AZ 85281, USA. He is now a research fellow at Nanyang Technological University in Singapore.
 Email: \texttt{junzhao@alumni.cmu.edu} }\\~}}


\maketitle




%




\begin{abstract}
 We present a comprehensive view of the relations among several privacy notions: differential privacy (DP)~\cite{Dwork2006}, Bayesian differential privacy (BDP)~\cite{yang2015bayesian}, semantic privacy (SP)~\cite{kasiviswanathan2014semantics}, and membership privacy (MP)~\cite{Li-CCS2013}. The results are organized into two parts. In part one, we extend the notion of \mbox{semantic} privacy (SP) to Bayesian semantic privacy (BSP) and show its essential equivalence with Bayesian differential privacy (BDP) in the quantitative sense. We prove the relations between BDP, BSP, and SP as follows: \mbox{$\epsilon$-BDP $\Longleftarrow$ $\big(\frac{1}{2}-\frac{1}{e^{\epsilon}+1}\big)$-BSP}, and \mbox{$\epsilon$-BDP $\Longrightarrow$ $(e^{2\epsilon}-1)$-BSP $\Longrightarrow$ $(e^{2\epsilon}-1)$-SP}. In addition, we obtain a minor result   \mbox{$\epsilon$-DP $\Longleftarrow$ $\big(\frac{1}{2}-\frac{1}{e^{\epsilon}+1}\big)$-SP}, which   improves the result of Kasiviswanathan and Smith~\cite{kasiviswanathan2014semantics} stating $\epsilon$-DP $\Longleftarrow$ \mbox{$\epsilon/6$-SP} for $\epsilon \leq 1.35$.  In part two, we establish the   relations between BDP and MP.
First,  $\epsilon$-BDP $\Longrightarrow$ $\epsilon$-MP.
Second, for a family of distributions that are downward scalable in the sense of Li \emph{et al}.~\cite{Li-CCS2013}, it is shown that \mbox{$\epsilon$-BDP $\Longleftarrow$ $\epsilon$-MP.}
\end{abstract}
%

%
%
%

%


\section{Introduction}

\textbf{Differential privacy (DP).} Differential privacy by Dwork~\textit{et~al.}~\cite{Dwork2006,dwork2006calibrating} is a robust privacy standard that has been successfully
applied to a range of data analysis tasks, since it provides a rigorous foundation for defining and
preserving privacy. Differential privacy has received considerable attention in the literature~\cite{andres2013geo,zhang2016privtree,blocki2016differentially,xiao2015protecting,loucost,shokri2015privacy,KamalikaChaudhuriAllerton17,KamalikaChaudhuriSIGMOD17,shokri2017membership,phan2017adaptive}. Apple has incorporated differential privacy into its mobile operating system iOS 10~\cite{tang2017privacy}.  Google has implemented a differentially private tool called
RAPPOR in the Chrome browser to collect information about clients~\cite{erlingsson2014rappor}.
 A randomized algorithm $Y$ satisfies $\epsilon$-differentially privacy
if for all adjacent databases $x$, $x'$ and any event $E$, it holds that
\mbox{$\pr[Y(x) \in E] \leq e^{\epsilon} \pr[Y(x') \in E],$} where $\pr[\cdot]$ denotes the probability throughout this paper.  Intuitively, under differential privacy, an adversary
given access to the output do not have much confidence to determine
whether it was sampled from the probability distribution
generated by the algorithm when the database is $x$ or when the database is $x'$.

\textbf{Bayesian differential privacy (BDP).}  Yang \textit{et al.} ~\cite{yang2015bayesian} introduce the notion of Bayesian differential privacy as follows. Bayesian differential privacy broadens the application scenarios of differential privacy when data records have dependencies.  For a database $x$ with $n$ tuples,  let $i \in \{1,2,\ldots,n\}$ be a tuple index in the
database and $\mathcal{S} \subseteq \{1,2,\ldots,n\}\setminus {i}$ be a tuple index set. An
adversary denoted by $A(i, \mathcal{S}) $    knows
the values of all tuples in $\mathcal{S}$ (denoted by $x_\mathcal{S}$)  and attempts
to attack the value of tuple $i$ (denoted by $x_i$). For a randomized perturbation mechanism $Y = \pr[y \in \mathcal{Y} \con x]$ on database $x$, the Bayesian differential privacy leakage (BDPL) of $Y$ with respect to the adversary  $A(i, \mathcal{S}) $ is $\texttt{BDPL}_A(Y) = \text{sup}_{x_i, x_i', x_\mathcal{S},  \mathcal{Y}} \ln \frac{\pr[y \in \mathcal{Y} \con x_i, x_{\mathcal{S}}]}{\pr[y \in \mathcal{Y} \con x_i', x_\mathcal{S}]} $. The mechanism $Y$ satisfies $\epsilon$-Bayesian differential privacy if $\text{sup}_A \texttt{BDPL}_A(Y) \leq \epsilon$.

\textbf{Semantic privacy (SP).}
 Kasiviswanathan and Smith \cite{kasiviswanathan2014semantics}
 propose a Bayesian formulation of semantic privacy, inspired by the following interpretation of differential privacy explained in~\cite{Dwork2006}: \emph{Regardless of external knowledge, an adversary
with access to the sanitized database draws the same conclusions whether or not any individual data is included in the original database}. The phrases ``external knowledge'' and ``drawing conclusions'' are formulated as follows in \cite{kasiviswanathan2014semantics}. The external knowledge is modeled by
  a prior probability distribution $b$ on $\mathcal{D}^n$, where $b$ is short for ``belief'', and databases are assumed to be vectors in $\mathcal{D}^n$ for some domain $\mathcal{D}$. Conclusions are captured via the corresponding posterior distribution: given a transcript
$y$, the adversary updates his belief $b$ about the database $x$ using Bayes' rule to obtain a
posterior $\overline{b}$: $
\overline{b}[x|y]  =\frac{\bp{Y(x)=y}b[x]}{\sum_{z}\bp{Y(z)=y}b[z]}.$

 For the database $x$, Kasiviswanathan and Smith \cite{kasiviswanathan2014semantics} further define $x_{-i}$ to be the same vector except   that the record at position $i$ has been replaced
by some fixed, default value $\perp$ in $\mathcal{D}$.

 Kasiviswanathan and Smith \cite{kasiviswanathan2014semantics}
define $n + 1$ related games, numbered $0$ through $n$. In Game $0$, the adversary interacts
with $Y(x)$. This is the interaction that actually takes place between the adversary and
the randomized mechanism $Y$. Hence, the distribution $\overline{b}_0$ is just the distribution $\overline{b}$ as defined
in (\ref{bbar-uncorrelated}); i.e.,
$
\overline{b}_0[x|y]  =\overline{b}[x|y] =\frac{\bp{Y(x)=y}b[x]}{\sum_{z}\bp{Y(z)=y}b[z]}.$

In Game $i$ (for $1 \leq i \leq n$), the adversary interacts with $Y(x_{-i})$. Game $i$ describes
the hypothetical scenario where person $i$'s record is not used. In Game $i$ (for $1 \leq i \leq n$), given a
transcript $y$, the adversary updates his belief $b$ about database $x$ again using Bayes' rule
to obtain a posterior $\overline{b}_i$ as follows:
$
\overline{b}_i[x|y] =\frac{\bp{Y(x_{-i})=y}b[x]}{\sum_{z}\bp{Y(z_{-i})=y}b[z]}.$

Given a
transcript $y$, Kasiviswanathan and Smith \cite{kasiviswanathan2014semantics} say that privacy has been breached if the adversary
would draw different conclusions about the world and, in particular, about a person $i$,
depending on whether or not $i$'s data was used. To this end, Kasiviswanathan and Smith \cite{kasiviswanathan2014semantics} formally define $\epsilon$-semantic privacy below, where the statistical difference $\texttt{SD}(X,Y)$ between random variables $X$ and $Y$
on the same discrete space $D$ is defined by~$\texttt{SD}(X,Y)=\max_{S\subseteq D}\big|\bp{X\in S}-\bp{Y\in S}\big|.$ A randomized mechanism $Y$ is said to be $\epsilon$-semantically private if for all belief distributions $b$ on $\mathcal{D}^n$, for all possible transcripts $y$,
and for all $i = 1, \ldots, n$, it holds that
$
\textnormal{\texttt{SD}}(\overline{b}_0[\cdot|y],
\,\overline{b}_i[\cdot|y]) \leq \epsilon.$

\textbf{Membership privacy (MP).} Li \textit{et al.}  \cite{Li-CCS2013} propose membership privacy (MP) in  consideration of the adversary's prior beliefs. Let
the adversary's prior beliefs about   the dataset be
  captured by a distribution $\mathcal{D}$.
 From the adversary's point of view, the dataset is a random
variable drawn according to the distribution $\mathcal{D}$. With $\overline{x_i}$ denoting the event that record $x_i$ is not in the database, Li~\emph{et~al}.~\cite{Li-CCS2013}
define membership privacy as follows.
A mechanism ${Y}$ achieves ${\epsilon}$-membership privacy under a family $\mathbb{D}$ of distributions, i.e., $\langle\mathbb{D},\epsilon\rangle$-MP, if and only if for any distribution $\mathcal{D}\in\mathbb{D}$ and for any record $x_i$, any possible set $\mathcal{Y}$ for the output, we have\footnote{$\langle\mathbb{D},\epsilon\rangle$-membership privacy actually corresponds to $\langle\mathbb{D},e^\epsilon\rangle$-positive membership privacy in \cite{Li-CCS2013}. Li \emph{et al}. \cite{Li-CCS2013} use $\gamma$ and $\gamma^{-1}$ instead of $e^{\epsilon}$ and $e^{-\epsilon}$ in (\ref{MembershipPrivacy1}) and (\ref{MembershipPrivacy2}) to define $\langle\mathbb{D},\gamma\rangle$-membership privacy. We use $e^{\epsilon}$ and $e^{-\epsilon}$ here for better comparison between membership privacy and Bayesian differential privacy. Also, by membership privacy, we mean positive membership privacy of \cite{Li-CCS2013}. We do not discuss negative membership privacy of \cite{Li-CCS2013}.}
$  \mathbb{P}_{\mathcal{D},Y}[x_i\con \mathcal{Y}]    \leq e^{\epsilon}\mathbb{P}_{\mathcal{D}}[x_i] $
and
$ \mathbb{P}_{\mathcal{D},Y}[\overline{x_i}\con \mathcal{Y}]    \geq  e^{-\epsilon}\mathbb{P}_{\mathcal{D}}[\overline{x_i}]. $

The rest of the paper is organized as follows. Section
\ref{sec:main:res} presents the results on the relations among several privacy notions: differential privacy (DP), Bayesian differential privacy (BDP), semantic privacy (SP), and membership privacy (MP). We elaborate their proofs in Sections \ref{secprofco}.  Section \ref{related} surveys related work, and
Section \ref{sec:Conclusion} concludes the paper.

\section{The Results} \label{sec:main:res}

    Kasiviswanathan and Smith \cite{kasiviswanathan2014semantics} introduce semantic privacy (SP) and show its essential equivalence with differential privacy (DP) in the quantitative sense (the notion of essential equivalence means $\epsilon$-DP $\Longleftarrow$ $f(\epsilon)$-SP and $\epsilon$-DP $\Longrightarrow$ $g(\epsilon)$-SP for some functions $f$ and $g$). We extend their notion to Bayesian semantic privacy (BSP) and show its essential equivalence with Bayesian differential privacy (BDP) also in the quantitative sense. We prove the relations between BDP, BSP, and SP as follows:
 \begin{itemize}
  \item[(i)] $\epsilon$-BDP $\Longleftarrow$ $\big(\frac{1}{2}-\frac{1}{e^{\epsilon}+1}\big)$-BSP.
  \item[(ii)]  $\epsilon$-BDP $\Longrightarrow$ $(e^{2\epsilon}-1)$-BSP $\Longrightarrow$ $(e^{2\epsilon}-1)$-SP. 
\end{itemize}
We prove results (i) and (ii) in Section \ref{sec1}, where we also obtain a minor result   \mbox{$\epsilon$-DP $\Longleftarrow$ $\big(\frac{1}{2}-\frac{1}{e^{\epsilon}+1}\big)$-SP}, which   improves the result of Kasiviswanathan and Smith~\cite{kasiviswanathan2014semantics} stating \mbox{$\epsilon$-DP $\Longleftarrow$ \mbox{$\epsilon/6$-SP}} for $\epsilon \leq 1.35$.

Li \emph{et al}. \cite{Li-CCS2013} propose membership privacy (MP), which is applicable to Bayesian
data, in contrast to DP. However, no general algorithm has been
proposed for this framework. We present the following  relations between BDP and MP:
 \begin{itemize}
  \item[(iii)]  $\epsilon$-BDP $\Longrightarrow$ $\epsilon$-MP.
  \item[(iv)] For a family of distributions that are downward scalable in the sense of Li \emph{et al}. \cite{Li-CCS2013}, $\epsilon$-BDP $\Longleftarrow$ $\epsilon$-MP (See \cite{Li-CCS2013} for the meaning of ``downward scalable'' distributions).
\end{itemize}
We prove results (iii) and (iv) in Section \ref{sec2}.

\section{Proofs} \label{secprofco}

\subsection{Relations between our Bayesian differential privacy and Kasiviswanathan and Smith's semantic privacy \cite{kasiviswanathan2014semantics}}\label{sec1}
\newcommand{\bpd}[1]{{\mathbb{P}_{\mathcal{D}}}\left[{#1}\right]}
\newcommand{\bpdy}[1]{{\mathbb{P}_{\mathcal{D},Y}}\left[{#1}\right]}

 We extend the work of Kasiviswanathan and Smith \cite{kasiviswanathan2014semantics} on semantic privacy to tackle the case of correlated tuples. Specifically, we will present {Bayesian} semantic privacy and prove that the notions of {Bayesian} differential privacy and {Bayesian} semantic privacy are essentially (i.e., quantitatively) equivalent (see Theorem \ref{thmmain} below). Our result resembles \cite[Theorem 2.2]{kasiviswanathan2014semantics}, which shows that   differential privacy and   semantic privacy are essentially equivalent.



\begin{thm} \label{thmmain}
$\epsilon$-Bayesian differential privacy implies $(e^{2\epsilon}-1)$-Bayesian semantic privacy, and is implied by $\big(\frac{1}{2}-\frac{1}{e^{\epsilon}+1}\big)$-Bayesian semantic privacy.
\end{thm}

 \begin{thm}[Improving the result of Kasiviswanathan and Smith~\cite{kasiviswanathan2014semantics}] \label{thmmain1}
$\epsilon$-Differential privacy implies $(e^{2\epsilon}-1)$-semantic privacy, and is implied by $\big(\frac{1}{2}-\frac{1}{e^{\epsilon}+1}\big)$-semantic privacy.
\end{thm}

Theorem \ref{thmmain} is one of our novel results.
The first part of Theorem~\ref{thmmain1} is obtained by
Kasiviswanathan and Smith~\cite{kasiviswanathan2014semantics}. The second part of Theorem~\ref{thmmain1} improves the corresponding result of Kasiviswanathan and Smith~\cite{kasiviswanathan2014semantics}, which states that $\epsilon$-differential privacy is implied by $\epsilon/6$-semantic privacy for $\epsilon \leq 1.35$. The improvement can be seen from $\frac{1}{2}-\frac{1}{e^{\epsilon}+1} > \epsilon/6$ for $\epsilon \leq 1.35$.

The rest of the discussion is organized as follows.
We review semantic privacy and define Bayesian semantic privacy in Section \ref{spdsp}. In Section \ref{recddp}, we recall Bayesian differential privacy. Finally, we prove the above Theorem \ref{thmmain} in Section \ref{secrel}. The proof of Theorem \ref{thmmain1} is similar to that of Theorem \ref{thmmain}.

\subsubsection{Reviewing semantic privacy and defining Bayesian semantic privacy} \label{spdsp}

In this section, we first review semantic privacy from Kasiviswanathan and Smith \cite{kasiviswanathan2014semantics}, before presenting Bayesian semantic privacy, which extends the notion of semantic privacy to address correlated tuples.

\textbf{A review of Kasiviswanathan and Smith \cite{kasiviswanathan2014semantics} for semantic privacy:}

 Kasiviswanathan and Smith \cite{kasiviswanathan2014semantics}
 propose a Bayesian formulation of semantic privacy, inspired by the following interpretation of differential privacy explained in \cite{Dwork2006}: \emph{Regardless of external knowledge, an adversary
with access to the sanitized database draws the same conclusions whether or not any individual data is included in the original database}. The phrases ``external knowledge'' and ``drawing conclusions'' are formulated as follows in \cite{kasiviswanathan2014semantics}. The external knowledge is modeled by
   a prior probability distribution $b$ on $\mathcal{D}^n$, where $b$ is short for ``belief,'' and databases are assumed to be vectors in $\mathcal{D}^n$ for some domain $\mathcal{D}$. Conclusions are captured via the corresponding posterior distribution: given a transcript
$y$, the adversary updates his belief $b$ about the database $x$ using Bayes' rule to obtain a
posterior $\overline{b}$:\footnote{For simplicity, only discrete probability distributions are discussed. The results can be readily extended to the continuous case.}
 \begin{align}
\overline{b}[x|y] & =\frac{\bp{Y(x)=y}b[x]}{\sum_{z}\bp{Y(z)=y}b[z]}.
\label{bbar-uncorrelated}
\end{align}

 For the database $x$, Kasiviswanathan and Smith \cite{kasiviswanathan2014semantics} further define $x_{-i}$ to be the same vector except that position $i$ has been replaced
by some fixed, default value in $\mathcal{D}$. Any valid value in $\mathcal{D}$ will do for the default value. In addition, the default value can be understood as a special value $\perp$ (e.g., ``no data''); see \cite[Page 3--Footnote 2]{kasiviswanathan2014semantics} for details. We will use $\perp$ whenever it is necessary to explicitly write out the default value.

 Kasiviswanathan and Smith \cite{kasiviswanathan2014semantics}
define $n + 1$ related games, numbered $0$ through $n$. In Game $0$, the adversary interacts
with $Y(x)$. This is the interaction that actually takes place between the adversary and
the randomized mechanism $Y$. Hence, the distribution $\overline{b}_0$ is just the distribution $\overline{b}$ as defined
in (\ref{bbar-uncorrelated}); i.e.,
\begin{align}
\overline{b}_0[x|y] & =\overline{b}[x|y] =\frac{\bp{Y(x)=y}b[x]}{\sum_{z}\bp{Y(z)=y}b[z]}.
\label{bbar0-uncorrelated}
\end{align}

In Game $i$ (for $1 \leq i \leq n$), the adversary interacts with $Y(x_{-i})$. Game $i$ describes
the hypothetical scenario where person $i$'s record is not used. In Game $i$ (for $1 \leq i \leq n$), given a
transcript $y$, the adversary updates his belief $b$ about database $x$ again using Bayes' rule
to obtain a posterior $\overline{b}_i$ as follows:
\begin{align}
\overline{b}_i[x|y] & =\frac{\bp{Y(x_{-i})=y}b[x]}{\sum_{z}\bp{Y(z_{-i})=y}b[z]}.
\label{bbari-uncorrelated}
\end{align}

Given a
transcript $y$, Kasiviswanathan and Smith \cite{kasiviswanathan2014semantics} say that privacy has been breached if the adversary
would draw different conclusions about the world and, in particular, about a person $i$,
depending on whether or not $i$'s data was used. To this end, Kasiviswanathan and Smith \cite{kasiviswanathan2014semantics} formally define $\epsilon$-semantic privacy below, where the statistical difference $\texttt{SD}(X,Y)$ between probability distributions (or random variables) $X$ and $Y$
on a discrete space $D$ is defined by $$\texttt{SD}(X,Y)=\max_{S\subseteq D}\big|\bp{X\in S}-\bp{Y\in S}\big|.$$

\begin{defn}[$\epsilon$-Semantical Privacy by {\cite[Definition 2.1]{kasiviswanathan2014semantics}}] \label{defSP}
A randomized mechanism $Y$ is said to be $\epsilon$-semantically private if for all belief distributions $b$ on $\mathcal{D}^n$, for all possible transcripts $y$,
and for all $i = 1, \ldots, n$:
\begin{align}
\textnormal{\texttt{SD}}(\overline{b}_0[\cdot|y],
\,\overline{b}_i[\cdot|y]) \leq \epsilon.\label{bbar0i-uncorrelated}
\end{align}
 \end{defn}

From (\ref{bbari-uncorrelated}) and (\ref{bbar0i-uncorrelated}), the above definition of $\epsilon$-semantic privacy requires the use of $x_{-i}$, where $x_{-i}$ is obtained after we replace position $i$ at $x$ by the default value $\perp$. If the tuples are correlated, changing position $i$ at $x$ might also result in changing other positions at $x$. Hence, $\epsilon$-semantic privacy may not work well under correlated tuples. Given this, we next extend $\epsilon$-semantic privacy to address correlated tuples and present $\epsilon$-Bayesian semantic privacy.

\textbf{Extending semantic privacy to Bayesian semantic privacy to address correlated tuples:}

As will become clear, our extension of semantic privacy to Bayesian semantic privacy  is similar to the extension of differential privacy to Bayesian differential privacy.

We let a statistical database be $[X_1,X_2, \ldots, X_n]$, where $X_j$ for each $j\in \{1,2,\ldots,n\}$ is a \emph{random variable}. We also let $\mathcal{N}$ be $\{1,2,\ldots,n\}$. Then we consider the databases $x$ and $z$ used in (\ref{bbar-uncorrelated})--(\ref{bbari-uncorrelated}) above to be
\begin{align}
x & = [X_1=x_1,X_2=x_2,
\ldots,X_n=x_n] = [X_j=x_{j}:j \in \mathcal{N}], \label{xfull}
\end{align}
and
\begin{align}
z & = [X_1=z_1,X_2=z_2,
\ldots,X_n=z_n] = [X_j=z_{j}:j \in \mathcal{N}]. \label{zfull}
\end{align}

When the data tuples are correlated,
the adversary may gain more advantage in inferring $x_i$ by using random variables $X_j|_{j\in \mathcal{S}}$'s instantiations $x_j|_{j\in \mathcal{S}}$, and random variables $X_j|_{j\in \mathcal{N}\setminus\{i\}\setminus\mathcal{S}}$ for computation instead of using instantiations $x_j|_{j\in \mathcal{N}\setminus\{i\}}$ only, where $\mathcal{S} \subseteq \mathcal{N}\setminus\{i\}$ (note that $\mathcal{S}$ can be an arbitrary subset of $ \mathcal{N}\setminus\{i\}$). For notation convenience, we define $x_{i+\mathcal{S}}$ and $z_{i+\mathcal{S}}$ by
\begin{align}
x_{i+\mathcal{S}} & = [X_i=x_i ,~X_j=x_{j}:j \in \mathcal{S},~X_j:j \in \mathcal{N}\setminus\{i\}\setminus\mathcal{S}], \label{xiplusS}
\end{align}
and
\begin{align}
z_{i+\mathcal{S}} & = [X_i=z_i ,~X_j=z_{j}:j \in \mathcal{S},~X_j:j \in \mathcal{N}\setminus\{i\}\setminus\mathcal{S}].\label{ziplusS}
\end{align}
From (\ref{xfull})--(\ref{ziplusS}), if $\mathcal{S} = \mathcal{N}\setminus\{i\}$, then $x_{i+\mathcal{S}}$ and $z_{i+\mathcal{S}}$ reduce to databases $x$ and $z$, respectively.

 Similar to the previous subsection, here we also let the adversary play $n + 1$ related games with the randomized mechanism $Y$, and define $\overline{b},\overline{b}_0,\overline{b}_i|_{i=1,\ldots,n}$ as detailed below. In Game $0$, the adversary interacts
with $Y(x_{i+\mathcal{S}})$.
We generalize $x$ and $z$ in (\ref{bbar0-uncorrelated}) to $x_{i+\mathcal{S}}$ and $z_{i+\mathcal{S}}$, so that (\ref{bbar0-uncorrelated}) becomes
\begin{align}
\overline{b}_0[x_{i+\mathcal{S}}|y] & =\overline{b}[x_{i+\mathcal{S}}|y] =\frac{\bp{Y(x_{i+\mathcal{S}})=y}b[x_{i+\mathcal{S}}]}{\sum_{z_{i+\mathcal{S}}}\bp{Y(z_{i+\mathcal{S}})=y}b[z_{i+\mathcal{S}}]}.
\label{bbar0-correlated}
\end{align}
For clarity, we explain the beliefs in (\ref{bbar0-correlated}). From (\ref{xiplusS}) and (\ref{ziplusS}), $b[x_{i+\mathcal{S}}]$ and $b[z_{i+\mathcal{S}}]$ in (\ref{bbar0-correlated}) are given by
\begin{align}
b[x_{i+\mathcal{S}}] & = b[X_i=x_i ,~X_j=x_{j}:j \in \mathcal{S},~X_j:j \in \mathcal{N}\setminus\{i\}\setminus\mathcal{S}] \nonumber \\ & = b[X_i=x_i ,~X_j=x_{j}:j \in \mathcal{S}], \label{bxiplusS}
\end{align}
and
\begin{align}
b[z_{i+\mathcal{S}}] & = b[X_i=z_i ,~X_j=z_{j}:j \in \mathcal{S},~X_j:j \in \mathcal{N}\setminus\{i\}\setminus\mathcal{S}] \nonumber \\  & =b [X_i=z_i ,~X_j=z_{j}:j \in \mathcal{S}].\label{bziplusS}
\end{align}
Similar to (\ref{bxiplusS}), from (\ref{xiplusS}), $\overline{b}_0[x_{i+\mathcal{S}}|y]$ is given by
\begin{align}
& \overline{b}_0[x_{i+\mathcal{S}}|y]  \nonumber \\& = \overline{b}_0[X_i=x_i ,~X_j=x_{j}:j \in \mathcal{S},~X_j:j \in \mathcal{N}\setminus\{i\}\setminus\mathcal{S}|y] \nonumber \\ & = \overline{b}_0[X_i=x_i ,~X_j=x_{j}:j \in \mathcal{S}|y]. \label{b0xiplusS}
\end{align}

In Game $i$ (for $1 \leq i \leq n$), we change position $i$ at $x_{i+\mathcal{S}}$ to the default value $\perp$ to obtain $x_{-i+\mathcal{S}}$ defined below; specifically, recalling $x_{i+\mathcal{S}}$ given by (\ref{xiplusS}), we set $x_{-i+\mathcal{S}}$ by
\begin{align}
x_{-i+\mathcal{S}} & = [X_i=\perp ,~X_j=x_{j}:j \in \mathcal{S},~X_j:j \in \mathcal{N}\setminus\{i\}\setminus\mathcal{S}].
\end{align}
Similarly, we change position $i$ at $z_{i+\mathcal{S}}$ to the default value $\perp$ to obtain $z_{-i+\mathcal{S}}$ defined below; specifically, recalling $z_{i+\mathcal{S}}$ given by (\ref{ziplusS}), we set $z_{-i+\mathcal{S}}$ by \begin{align}
z_{-i+\mathcal{S}} & = [X_i=\perp ,~X_j=z_{j}:j \in \mathcal{S},~X_j:j \in \mathcal{N}\setminus\{i\}\setminus\mathcal{S}].
\end{align}
As $x_{i+\mathcal{S}}$ and $z_{i+\mathcal{S}}$  generalize $x$ and $z$ in (\ref{bbari-uncorrelated}), clearly $x_{-i+\mathcal{S}}$ and $z_{-i+\mathcal{S}}$ also generalize  $x_{-i}$ and $z_{-i}$ in (\ref{bbari-uncorrelated}). In Game $i$ (for $1 \leq i \leq n$), the adversary interacts with $Y(x_{-i+\mathcal{S}})$. Then replacing $x$, $z$, $x_{-i}$ and $z_{-i}$ in (\ref{bbari-uncorrelated}) by $x_{i+\mathcal{S}}$, $z_{i+\mathcal{S}}$, $x_{-i+\mathcal{S}}$ and $z_{-i+\mathcal{S}}$, respectively, we obtain
\begin{align}
\overline{b}_i[x_{i+\mathcal{S}}|y] & =\frac{\bp{Y(x_{-i+\mathcal{S}})=y}b[x_{i+\mathcal{S}}]}{\sum_{z_{i+\mathcal{S}}}\bp{Y(z_{-i+\mathcal{S}})=y}b[z_{i+\mathcal{S}}]}.
\label{bi}
\end{align}
The beliefs $b[x_{i+\mathcal{S}}]$ and $b[z_{i+\mathcal{S}}]$ in (\ref{bi}) are already interpreted as (\ref{bxiplusS}) and (\ref{bziplusS}).
For clarity, we further explain $\overline{b}_i[x_{i+\mathcal{S}}|y]$ in (\ref{bi}).
Similar to (\ref{b0xiplusS}), from (\ref{xiplusS}), $\overline{b}_i[x_{i+\mathcal{S}}|y]$ is given by
\begin{align}
&\overline{b}_i[x_{i+\mathcal{S}}|y] \nonumber \\ & = \overline{b}_i[X_i=x_i ,~X_j=x_{j}:j \in \mathcal{S},~X_j:j \in \mathcal{N}\setminus\{i\}\setminus\mathcal{S}|y] \nonumber \\ & = \overline{b}_i[X_i=x_i ,~X_j=x_{j}:j \in \mathcal{S}|y]. \label{bixiplusS}
\end{align}

With the above notation, we define $\epsilon$-Bayesian semantical privacy below, in a way similar to that of $\epsilon$-semantical privacy in Definition \ref{defSP}.

\begin{defn}[$\epsilon$-Bayesian Semantical Privacy] \label{defBSP}
A randomized mechanism $Y$ is said to have $\epsilon$-Bayesian semantical privacy if for all belief distributions $b$ on $\mathcal{D}^n$, for all possible transcripts $y$,
 for all $i = 1, \ldots, n$, and for all $x_{i+\mathcal{S}}$ and $z_{i+\mathcal{S}}$ defined in (\ref{xiplusS}) and (\ref{ziplusS}) with $\mathcal{S}\subseteq \mathcal{N}\setminus\{i\}$:
\begin{align}
\textnormal{\texttt{SD}}(\overline{b}_0[x_{i+\mathcal{S}}|y],
\,\overline{b}_i[x_{i+\mathcal{S}}|y]) \leq \epsilon. \label{bbar0i-correlated}
\end{align}
 \end{defn}

To understand the beliefs $\overline{b}_0[x_{i+\mathcal{S}}|y]$ and $\overline{b}_i[x_{i+\mathcal{S}}|y]$ in  (\ref{bbar0i-correlated}), we use their interpretations in (\ref{b0xiplusS}) and  (\ref{bixiplusS}). In Definition \ref{defBSP} for $\epsilon$-Bayesian semantical privacy, we consider all possible $\mathcal{S}\subseteq \mathcal{N}\setminus\{i\}$. In the hypothetical scenario where we  consider $\mathcal{S}$ only as $ \mathcal{N}\setminus\{i\}$ in Definition \ref{defBSP}, Definition \ref{defBSP} would reduce to Definition \ref{defSP} for $\epsilon$-semantical privacy.

\subsubsection{Recalling Bayesian differential privacy} \label{recddp}

In this section, we recall Bayesian differential privacy and express its definition using some new notation.

With $x_{i+\mathcal{S}}$ defined in (\ref{xiplusS}) (i.e., $x_{i+\mathcal{S}}  = [X_i=x_i ,~X_j=x_{j}:j \in \mathcal{S},~X_j:j \in \mathcal{N}\setminus\{i\}\setminus\mathcal{S}]$), for notation convenience, we further define
 $x_{i+\mathcal{S}}'$ by
\begin{align}
x_{i+\mathcal{S}}'  = [X_i=x_i' ,~X_j=x_{j}:j \in \mathcal{S},~X_j:j \in \mathcal{N}\setminus\{i\}\setminus\mathcal{S}]. \label{xiplusSpr}
\end{align}
Note that the only difference between $x_{i+\mathcal{S}}$ and $x_{i+\mathcal{S}}'$ is that the former has $X_i=x_i$, while the latter enforces $X_i=x_i'$. Then  $\epsilon$-Bayesian differential privacy means \begin{align}
\frac{\bp{Y(x_{i+\mathcal{S}})=y}}{\bp{Y(x_{i+\mathcal{S}}')=y}}
 \leq e^{\epsilon}. \label{ddprecall}
\end{align}

\subsubsection{Proving Theorem \ref{thmmain} on the relations between Bayesian differential privacy and Bayesian semantic privacy} \label{secrel}

Our Theorem~\ref{thmmain} restated below presents the relations between Bayesian differential privacy and Bayesian semantic privacy.\\

\noindent \textbf{Theorem \ref{thmmain} (Restated).~}
{\em $\epsilon$-Bayesian differential privacy implies $(e^{2\epsilon}-1)$-Bayesian semantic privacy, and is implied by $\big(\frac{1}{2}-\frac{1}{e^{\epsilon}+1}\big)$-Bayesian semantic privacy.\\
}

Theorem \ref{thmmain} shows that the notions of Bayesian differential privacy and Bayesian semantic privacy are essentially equivalent (of course, the parameters should be set appropriately). The proof of Theorem \ref{thmmain} below is just an extension of the reasoning by Kasiviswanathan and Smith \cite{kasiviswanathan2014semantics}.\\

\noindent \textbf{Proof of Theorem \ref{thmmain}.~} We show Theorem \ref{thmmain} in two parts below. We will use the following definition of point-wise $(\epsilon, 0)$-indistinguishability from \cite[Definition 3.2]{kasiviswanathan2014semantics}: Two discrete random variables $X$
and $Y$ are point-wise $(\epsilon, 0)$-indistinguishable if it holds for a drawn
from either $X$
or $Y$ that $e^{-\epsilon}\bp{Y=a}\leq  \bp{X=a} \leq e^{\epsilon}\bp{Y=a}$.

 \emph{Proving $\epsilon$-Bayesian differential privacy $\Longrightarrow$ $(e^{2\epsilon}-1)$-Bayesian semantic privacy:~}
To prove this part, we consider any database $x \in \mathcal{D}^n$. Let $Y$ be an $\epsilon/2$-Bayesian differentially private algorithm.
Consider any belief distribution $b$. Let the posterior distributions $\overline{b}_0[x_{i+\mathcal{S}}|y]$ and $\overline{b}_i[x_{i+\mathcal{S}}|y]$ for
some fixed $i$, $\mathcal{S}$ and $y$ be defined in (\ref{bbar0-correlated}) and (\ref{bi}). From (\ref{ddprecall}), $\epsilon$-Bayesian differential privacy implies that for
every   $z_{i+\mathcal{S}}$,
\begin{align}
 e^{-\epsilon} \bp{Y(z_{-i+\mathcal{S}})=y}\leq \bp{Y(z_{i+\mathcal{S}})=y} \leq e^{\epsilon} \bp{Y(z_{-i+\mathcal{S}})=y}.\nonumber
\end{align}
These inequalities imply that the ratio of $\overline{b}_0[x_{i+\mathcal{S}}|y]$ and $\overline{b}_i[x_{i+\mathcal{S}}|y]$ (defined in (\ref{bbar0-correlated}) and (\ref{bi})) is
within $e^{\pm2\epsilon}$. Since these inequalities hold for every $x_{i+\mathcal{S}}$, we get:
\begin{align}
 e^{-2\epsilon}\overline{b}_i[x_{i+\mathcal{S}}|y]\leq \overline{b}_0[x_{i+\mathcal{S}}|y] \leq e^{2\epsilon}\overline{b}_i[x_{i+\mathcal{S}}|y],
 ~\forall x_{i+\mathcal{S}}\nonumber.
\end{align}
This implies that the random variables $\overline{b}_0[x_{i+\mathcal{S}}|y]$ and $\overline{b}_i[x_{i+\mathcal{S}}|y]$ are point-wise $(2\epsilon, 0)$-indistinguishable.
Applying \cite[Lemma 3.3-Property 5]{kasiviswanathan2014semantics}, we obtain $\textrm{\texttt{SD}}(\overline{b}_0[x_{i+\mathcal{S}}|y],
\,\overline{b}_i[x_{i+\mathcal{S}}|y]) \leq (e^{2\epsilon}-1)$. Repeating
the above arguments for every belief distribution, for every $i$, and for every $y$,
we thus show that the mechanism $Y$ is $(e^{2\epsilon}-1)$-Bayesian semantic private.

\emph{Proving $\big(\frac{1}{2}-\frac{1}{e^{\epsilon}+1}\big)$-Bayesian semantic privacy $\Longrightarrow$ \mbox{$\epsilon$-Bayesian differential privacy}:~}To prove this part, we consider a belief distribution
$b$ which is uniform over $$x_{i+\mathcal{S}}  = [X_i=x_i ,~X_j=x_{j}:j \in \mathcal{S},~X_j:j \in \mathcal{N}\setminus\{i\}\setminus\mathcal{S}]$$ and $$x_{i+\mathcal{S}}'  = [X_i=x_i' ,~X_j=x_{j}:j \in \mathcal{S},~X_j:j \in \mathcal{N}\setminus\{i\}\setminus\mathcal{S}];$$ i.e.,
\begin{align}
b[x_{i+\mathcal{S}}] & = b[X_i=x_i ,~X_j=x_{j}:j \in \mathcal{S}] = \frac{1}{2}\nonumber
\end{align}
and
\begin{align}
b[x_{i+\mathcal{S}}'] & = b[X_i=x_i' ,~X_j=x_{j}:j \in \mathcal{S}] = \frac{1}{2}.\nonumber
\end{align}
Fix a transcript $y$. The distribution
$\overline{b}_i[\cdot|y]$ will be uniform over $x_{i+\mathcal{S}}$ and $x_{i+\mathcal{S}}'$ since they induce the same distribution on
transcripts in Game $i$. This means that $\overline{b}_0[\cdot|y]$ will assign probabilities in the interval $[\frac{1}{2}-\big(\frac{1}{2}-\frac{1}{e^{\epsilon}+1}\big),\frac{1}{2}+\big(\frac{1}{2}-\frac{1}{e^{\epsilon}+1}\big)]$ to each
of $x_{i+\mathcal{S}}$ and $x_{i+\mathcal{S}}'$ (by Definition \ref{defSP}). Working through Bayes' rule shows that (note
that $b[x_{i+\mathcal{S}}] = b[x_{i+\mathcal{S}}']$)
\begin{align}
&\frac{\bp{Y(x_{i+\mathcal{S}})=y}}{\bp{Y(x_{i+\mathcal{S}}')=y}}
 \nonumber \\ &=\frac{\overline{b}_0[x_{i+\mathcal{S}}|y]}{\overline{b}_0[x_{i+\mathcal{S}}'|y]}
\leq \frac{\frac{1}{2}+\big(\frac{1}{2}-\frac{1}{e^{\epsilon}+1}\big)}{\frac{1}{2}-\big(\frac{1}{2}-\frac{1}{e^{\epsilon}+1}\big)} = e^{\epsilon}. \label{boundd}
\end{align}
Since the bound in (\ref{boundd}) holds for every $y$, $Y(x_{i+\mathcal{S}})$ and $Y(x_{i+\mathcal{S}}')$ are point-wise $(\epsilon, 0)$-indistinguishable.
From \cite[Lemma 3.3-Property 5]{kasiviswanathan2014semantics},  $Y(x_{i+\mathcal{S}})$ and $Y(x_{i+\mathcal{S}}')$ are $(\epsilon, 0)$-indistinguishable.
Since this relation holds for every pair of $x_{i+\mathcal{S}}$ and $x_{i+\mathcal{S}}'$,
the mechanism $Y$ is $\epsilon$-Bayesian differentially private. \pfe

\subsection{Relations between Bayesian differential privacy and membership privacy} \label{sec2}
The adversary may have prior beliefs about what the dataset is;
this is captured by a distribution $\mathcal{D}$.
 From the adversary's point of view, the dataset is a random
variable drawn according to the distribution $\mathcal{D}$. With $\overline{x_i}$ denoting the event that record $x_i$ is not in the database, Li \emph{et al}. \cite{Li-CCS2013}
define membership privacy as follows, where we reuse some notation of Li \emph{et al}. \cite{Li-CCS2013}.

\begin{defn}[\hspace{-.1pt}Li \emph{et al}. \cite{Li-CCS2013}\hspace{-.2pt}] \label{MPdef}
A mechanism ${Y}$ achieves \mbox{${\epsilon}$-membership privacy} under a family $\mathbb{D}$ of distributions, i.e., \mbox{$\langle\mathbb{D},\epsilon\rangle$-membership privacy}, if and only if for any distribution $\mathcal{D}\in\mathbb{D}$ and for any record $x_i$, any possible set $\mathcal{Y}$ for the output, we have\footnote{$\langle\mathbb{D},\epsilon\rangle$-membership privacy actually corresponds to $\langle\mathbb{D},e^\epsilon\rangle$-membership privacy in \cite{Li-CCS2013}. Li \emph{et al}. \cite{Li-CCS2013} use $\gamma$ and $\gamma^{-1}$ instead of $e^{\epsilon}$ and $e^{-\epsilon}$ in (\ref{MembershipPrivacy1}) and (\ref{MembershipPrivacy2}) to define $\langle\mathbb{D},\gamma\rangle$-membership privacy. We use $e^{\epsilon}$ and $e^{-\epsilon}$ here for better comparison between membership privacy and Bayesian differential privacy.}
\begin{align}
  \mathbb{P}_{\mathcal{D},Y}[x_i\con \mathcal{Y}]   &  \leq e^{\epsilon}\mathbb{P}_{\mathcal{D}}[x_i]  \label{MembershipPrivacy1}
\end{align}
and
\begin{align}
 \mathbb{P}_{\mathcal{D},Y}[\overline{x_i}\con \mathcal{Y}]   & \geq  e^{-\epsilon}\mathbb{P}_{\mathcal{D}}[\overline{x_i}]. \label{MembershipPrivacy2}
\end{align}
\end{defn}

We discuss the adversary model considered here.
Let $\mathcal{D}_{i,K}$ denote a distribution where $\bp{x_i,x_K}=p$ and $\bp{x_i',x_K}=1-p$ for some $p$. Define
$\mathbb{D}_*\de\cup_{\begin{subarray}{l}i\in \{1,\iffalse 2,\fi \ldots,n\},\\K\subseteq \{1,\iffalse 2,\fi \ldots,n\}\setminus\{i\}\end{subarray}}\mathcal{D}_{i,K}$. The adversary model will be captured by the family $\mathbb{D}_*$ of distributions.
For simplicity, we will refer to $\langle\mathbb{D}_*,\epsilon\rangle$-membership privacy as $\epsilon$-membership privacy.

\subsubsection{From Bayesian differential privacy to membership privacy}

\begin{thm} \label{lemBDPtoMP}
$\epsilon$-Bayesian differential privacy implies $\epsilon$-membership privacy.
\end{thm}

\begin{lem} \label{lemfinegrained}

A mechanism ${Y}$ achieves ${\epsilon}$-membership privacy under a family $\mathbb{D}$ of distributions, i.e., $\langle\mathbb{D},\epsilon\rangle$-MP, if and only if it holds for any distribution $\mathcal{D}\in\mathbb{D}$ that\footnote{We let $\frac{0}{0}=1$ and $\frac{\textrm{non-zero}}{0}=\infty$ to address the degenerate cases.}
\begin{subnumcases}{\hspace{-19pt}\frac{\mathbb{P}_{\mathcal{D},Y}[\mathcal{Y}\con x_i]}{\mathbb{P}_{\mathcal{D},Y}[\mathcal{Y}\con \overline{x_i}]}  \leq\hspace{-2pt}}\hspace{-2pt}
\frac{1-\mathbb{P}_{\mathcal{D}}[x_i]}{e^{-\epsilon}-\mathbb{P}_{\mathcal{D}}[x_i]},&\hspace{-15pt}\textrm{if $0 \leq \mathbb{P}_{\mathcal{D}}[x_i] \leq \frac{1}{1+e^{\epsilon}}$,}  \label{functionfexp1}\vspace{5pt}\\ \hspace{-2pt}
\frac{e^{\epsilon}-1+\mathbb{P}_{\mathcal{D}}[x_i]}{\mathbb{P}_{\mathcal{D}}[x_i]},&\hspace{-15pt}\textrm{if $\frac{1}{1+e^{\epsilon}} < \mathbb{P}_{\mathcal{D}}[x_i] \leq 1$.} \label{functionfexp2}
\end{subnumcases}
\end{lem}

We will explain that Lemma \ref{lemfinegrained} implies the following corollary, which will be used to show Theorem~\ref{lemBDPtoMP}.

\begin{cor} \label{corfinegrained}

A mechanism ${Y}$ achieves ${\epsilon}$-membership privacy under a family $\mathbb{D}$ of distributions, i.e., $\langle\mathbb{D},\epsilon\rangle$-MP, if it holds for any distribution $\mathcal{D}\in\mathbb{D}$ that $\frac{\mathbb{P}_{\mathcal{D},Y}[\mathcal{Y}\con x_i]}{\mathbb{P}_{\mathcal{D},Y}[\mathcal{Y}\con \overline{x_i}]}  \leq e^{\epsilon}$.

\end{cor}

\noindent \textbf{Proof of Theorem~\ref{lemBDPtoMP} using Corollary \ref{corfinegrained}.}
Under distribution $\mathcal{D}_{i,K}$ where $\bp{x_i,x_K}=p$ and $\bp{x_i',x_K}=1-p$ for some $p$, we have
\begin{align}
\mathbb{P}_{\mathcal{D}_{i,K},Y}[\mathcal{Y}\con x_i] & =  \bigp{Y(x_i,x_K,X_{\overline{K}})\in \mathcal{Y}}, \label{eqpdyxi1}
\end{align}
and
\begin{align}
\mathbb{P}_{\mathcal{D}_{i,K},Y}[\mathcal{Y}\con \overline{x_i}] & =  \bigp{Y(x_i',x_K,X_{\overline{K}})\in \mathcal{Y}}.\label{eqpdyxi2}
\end{align}
Under $\epsilon$-Bayesian differential privacy, we have
$\frac{\bigp{Y(x_i,x_K,X_{\overline{K}})\in \mathcal{Y}}}{\bigp{Y(x_i',x_K,X_{\overline{K}})\in \mathcal{Y}}}$\\$ \leq e^{\epsilon}$, which along with (\ref{eqpdyxi1}) and (\ref{eqpdyxi2}) yields $\frac{\mathbb{P}_{\mathcal{D}_{i,K},Y}[\mathcal{Y}\con x_i]}{\mathbb{P}_{\mathcal{D}_{i,K},Y}[\mathcal{Y}\con \overline{x_i}]}  \leq e^{\epsilon}$, then $\langle\mathbb{D}_*,\epsilon\rangle$-membership privacy (i.e., $\epsilon$-membership privacy) follows for
$\mathbb{D}_*\de\cup_{\begin{subarray}{l}i\in \{1,\iffalse 2,\fi \ldots,n\},\\K\subseteq \{1,\iffalse 2,\fi \ldots,n\}\setminus\{i\}\end{subarray}}\mathcal{D}_{i,K}$. \pfe

\noindent \textbf{Proof of Corollary \ref{corfinegrained} using Lemma \ref{lemfinegrained}.} Note that
(\ref{functionfexp1}) and  (\ref{functionfexp2}) in  Lemma \ref{lemfinegrained} can be written as $\frac{\mathbb{P}_{\mathcal{D},Y}[\mathcal{Y}\con x_i]}{\mathbb{P}_{\mathcal{D},Y}[\mathcal{Y}\con \overline{x_i}]} \leq g\big(\mathbb{P}_{\mathcal{D}}[x_i]\big)$, where
$g(b)$ is a function defined as follows:
\begin{align}
 g(b)  &  \de
\begin{cases}
\frac{1-b}{e^{-\epsilon}-b},&\textrm{if $0 \leq b \leq \frac{1}{1+e^{\epsilon}}$,} \vspace{5pt}\\
\frac{e^{\epsilon}-1+b}{b},&\textrm{if $\frac{1}{1+e^{\epsilon}} < b \leq 1$.} \label{functiongexpr}
\end{cases}
\end{align}
The function $g(b)$ increases as $b$ increases for $0 \leq b \leq \frac{1}{1+e^{\epsilon}}$ and  decreases as $b$ increases for $\frac{1}{1+e^{\epsilon}} < b \leq 1$. Hence, at $  b = 0$ or $b=1$, $g(b)$ takes its minimum $g(0)=g(1)=e^{\epsilon}$. Then $\frac{\mathbb{P}_{\mathcal{D},Y}[\mathcal{Y}\con x_i]}{\mathbb{P}_{\mathcal{D},Y}[\mathcal{Y}\con \overline{x_i}]} \leq e^{\epsilon}$ implies  $\frac{\mathbb{P}_{\mathcal{D},Y}[\mathcal{Y}\con x_i]}{\mathbb{P}_{\mathcal{D},Y}[\mathcal{Y}\con \overline{x_i}]} \leq g\big(\mathbb{P}_{\mathcal{D}}[x_i]\big)$ for any $\mathbb{P}_{\mathcal{D}}[x_i]$. In view this, we obtain Corollary \ref{corfinegrained} from   Lemma \ref{lemfinegrained}.


\noindent
\textbf{Proof of Lemma \ref{lemfinegrained}.} 
 For simplicity, we define
\begin{align}
A &\de \frac{\bpdy{\mathcal{Y}\boldsymbol{\mid}
 x_i}}{\bpdy{\mathcal{Y}\boldsymbol{\mid}
 \overline{x_i}}}, \label{equiva2defmn}
\end{align}

Then the goal of Lemma \ref{lemfinegrained} is to show the combination of (\ref{MembershipPrivacy1}) and (\ref{MembershipPrivacy2}) is equivalent to $A \leq g\big(\mathbb{P}_{\mathcal{D}}[x_i]\big)$.
Hence, we will establish Lemma \ref{lemfinegrained} once proving the following three results:
\begin{align}
(\ref{MembershipPrivacy1})& \Longleftrightarrow \Big\{ 1- \mathbb{P}_{\mathcal{D}}[x_i] \geq A \big(e^{-\epsilon} - \mathbb{P}_{\mathcal{D}}[x_i]\big) \Big\}, \label{equiva1}\\
(\ref{MembershipPrivacy2})& \Longleftrightarrow \Big\{A \times \mathbb{P}_{\mathcal{D}}[x_i] + 1 - \mathbb{P}_{\mathcal{D}}[x_i]   \leq e^{\epsilon}\Big\}, \label{equiva2}
\end{align}
and
\begin{align}
&\begin{rcases}
  1- \mathbb{P}_{\mathcal{D}}[x_i] \geq A \big(e^{-\epsilon} - \mathbb{P}_{\mathcal{D}}[x_i]\big), \\
 \text{and }A \times \mathbb{P}_{\mathcal{D}}[x_i] + 1 - \mathbb{P}_{\mathcal{D}}[x_i]   \leq e^{\epsilon}
\end{rcases}  \Longleftrightarrow A \leq g\big(\mathbb{P}_{\mathcal{D}}[x_i]\big).
\label{equiva3}
\end{align}
Below we demonstrate (\ref{equiva1}) (\ref{equiva2}) and  (\ref{equiva3}), respectively.

 \textbf{Proving (\ref{equiva1}):}

By Bayes' theorem, it holds that
\begin{align}
 &
\bpdy{ x_i\boldsymbol{\mid}
\mathcal{Y}} =  \frac{\bpdy{\mathcal{Y}\boldsymbol{\mid}
 x_i}\mathbb{P}_{\mathcal{D}}[x_i]}{\bpdy{\mathcal{Y}}}.\label{psq1Bayesianprivacya}
\end{align}
Given (\ref{psq1Bayesianprivacya}), we have
\begin{align}
 &(\ref{MembershipPrivacy1}) \Longleftrightarrow {\bpdy{\mathcal{Y}\boldsymbol{\mid}
 x_i} }  \leq e^{\epsilon}\times {\bpdy{\mathcal{Y}}}
.\label{psq1Bayesianprivacyb}
\end{align}
To prove (\ref{psq1Bayesianprivacyb}), we express ${\bpdy{\mathcal{Y}}}$ by the law of total probability, and find
\begin{align}
 &\bpdy{\mathcal{Y} }
= \bpdy{\mathcal{Y}\boldsymbol{\mid}
 x_i}\mathbb{P}_{\mathcal{D}}[x_i] +
 \bpdy{\mathcal{Y}\boldsymbol{\mid}
 \overline{x_i}}\mathbb{P}_{\mathcal{D}}[\overline{x_i}] 
 .\label{psq1Bayesianprivacyc}
\end{align}
Applying (\ref{equiva2defmn}) to  (\ref{psq1Bayesianprivacyc}), we obtain
\begin{align}
 &\bpdy{\mathcal{Y} }= \bpdy{\mathcal{Y}\boldsymbol{\mid}
 x_i}  \times  \Big\{\mathbb{P}_{\mathcal{D}}[x_i]
 +  A^{-1}\times
 \mathbb{P}_{\mathcal{D}}[\overline{x_i}]\Big\} .
 \label{psq1Bayesianprivacy3alignz} \end{align}
Then it follows from (\ref{psq1Bayesianprivacyb}) and (\ref{psq1Bayesianprivacy3alignz}) that
\begin{align}
(\ref{MembershipPrivacy1}) &\Longleftrightarrow \mathbb{P}_{\mathcal{D}}[x_i]
 +  A^{-1}\times
 \mathbb{P}_{\mathcal{D}}[\overline{x_i}]  \geq e^{-\epsilon}
 \nonumber \\ & \Longleftrightarrow
 1- \mathbb{P}_{\mathcal{D}}[x_i] \geq A \big(e^{-\epsilon} - \mathbb{P}_{\mathcal{D}}[x_i]\big); \nonumber
\end{align}
i.e., (\ref{equiva1}) is established.

 \textbf{Proving (\ref{equiva2}):}

By Bayes' theorem, it holds that
\begin{align}
 &
\bpdy{ \overline{x_i}\boldsymbol{\mid}
\mathcal{Y}} =  \frac{\bpdy{\mathcal{Y}\boldsymbol{\mid}
 \overline{x_i}}\mathbb{P}_{\mathcal{D}}[\overline{x_i}]}{\bpdy{\mathcal{Y}}}.\label{part2psq1Bayesianprivacya}
\end{align}
Given (\ref{part2psq1Bayesianprivacya}), we have
\begin{align}
 (\ref{MembershipPrivacy2}) \Longleftrightarrow&{\bpdy{\mathcal{Y}\boldsymbol{\mid}
 \overline{x_i}} }  \geq e^{-\epsilon}\times {\bpdy{Y(\boldsymbol{\mathscr{X}})\neq\boldsymbol{y}}}
.\label{part2psq1Bayesianprivacybs}
\end{align}
We recall (\ref{psq1Bayesianprivacy3alignz}). Applying (\ref{equiva2defmn}) to  (\ref{psq1Bayesianprivacy3alignz}), we obtain
\begin{align}
& \bpdy{\mathcal{Y} }\nonumber \\ & = A \times \bpdy{\mathcal{Y}\boldsymbol{\mid}
 \overline{x_i}}  \times  \Big\{\mathbb{P}_{\mathcal{D}}[x_i]
 +  A^{-1}\times
 \mathbb{P}_{\mathcal{D}}[\overline{x_i}]\Big\} .
 \label{psq1Bayesianprivacy3alignzpt2} \end{align}
Then it follows from (\ref{part2psq1Bayesianprivacybs}) and (\ref{psq1Bayesianprivacy3alignzpt2}) that
\begin{align}
(\ref{MembershipPrivacy2}) &\Longleftrightarrow A \times \Big\{\mathbb{P}_{\mathcal{D}}[x_i]
 +  A^{-1}\times
 \bpd{\overline{x_i}}\Big\}  \leq e^{\epsilon}
 \nonumber \\ & \Longleftrightarrow
 A \times\mathbb{P}_{\mathcal{D}}[x_i] + 1 - \mathbb{P}_{\mathcal{D}}[x_i]   \leq e^{\epsilon}  ; \nonumber
\end{align}
i.e., (\ref{equiva2}) is established.

 \textbf{Proving (\ref{equiva3}):}

With $\mathbb{P}_{\mathcal{D}}[x_i]$ replaced by real $x \in [0,1]$,  (\ref{equiva3}) will follow once we show for $x \in [0,1]$ that
\begin{align}
 \begin{rcases}
  1- x \geq A \big(e^{-\epsilon} - x\big), \\
 \text{and }A \times x + 1 - x   \leq e^{\epsilon}
\end{rcases}  & \Longleftrightarrow A \leq g(x).
\label{equiva3x}
\end{align}

We first prove the ``$\Longrightarrow$'' part in (\ref{equiva3x}). If $0 \leq x < e^{-\epsilon}$, we obtain from $1- x \geq A \big(e^{-\epsilon} - x\big)$ that $A \leq \frac{1-x}{e^{-\epsilon}-x}$. If $0<x \leq 1$, we obtain from $A \times x + 1 - x   \leq e^{\epsilon}$ that $A \leq \frac{e^{\epsilon}-1+x}{x}$. With $g_1(x)$ denoting $\frac{1-x}{e^{-\epsilon}-x}$ for $0 \leq x < e^{-\epsilon}$ and $g_2(x)$ denoting $\frac{e^{\epsilon}-1+x}{x}$ for $0<x \leq 1$, we see that $g(x)$ equals $g_1(x)$ if $0 \leq x \leq \frac{1}{1+e^{\epsilon}}$, and equals $g_2(x)$  if $\frac{1}{1+e^{\epsilon}} < x \leq 1$. Given the above, if $0 \leq x \leq \frac{1}{1+e^{\epsilon}}$, we have $A \leq g_1(x) = g(x)$, and if $\frac{1}{1+e^{\epsilon}} < x \leq 1$, we have $A \leq g_2(x) = g(x)$. Hence, the ``$\Longrightarrow$'' part in (\ref{equiva3x}) immediately follows.

We then prove the ``$\Longleftarrow$'' part in (\ref{equiva3x}). For any $x \in [0,1]$, we will establish i) $1- x \geq A \big(e^{-\epsilon} - x\big)$, and ii) $A \times x + 1 - x   \leq e^{\epsilon}$, respectively. We still use $g_1(x)$ and $g_2(x)$ defined above. Note that $g_1(x)$ is only defined for $0 \leq x < e^{-\epsilon}$ and $g_2(x)$ is only defined for $0<x \leq 1$. It is straightforward to show $g_1(x)\leq g_2(x)$ if $0 < x \leq \frac{1}{1+e^{\epsilon}}$, and $g_1(x)\geq g_2(x)$ if $\frac{1}{1+e^{\epsilon}} < x < e^{-\epsilon}$.
\begin{itemize}
  \item[i)] If $0 \leq x \leq \frac{1}{1+e^{\epsilon}}$, we obtain from $A \leq g(x) = g_1(x)$ that $A \leq \frac{1-x}{e^{-\epsilon}-x}$, implying $1- x \geq A \big(e^{-\epsilon} - x\big)$. If $\frac{1}{1+e^{\epsilon}} < x < e^{-\epsilon} $, we obtain from $A \leq g(x) = g_2(x)\leq g_1(x)$ that $A \leq \frac{1-x}{e^{-\epsilon}-x}$, yielding $1- x \geq A \big(e^{-\epsilon} - x\big)$. If $e^{-\epsilon} \leq x \leq 1$, it holds that $1- x \geq 0 \geq A \big(e^{-\epsilon} - x\big)$. To summarize, for any $x \in [0,1]$, it follows that $1- x \geq A \big(e^{-\epsilon} - x\big)$.
  \item[ii)] If $\frac{1}{1+e^{\epsilon}} < x \leq 1$, we obtain from $A \leq g(x) = g_2(x)$ that $A \leq \frac{e^{\epsilon}-1+x}{x}$, implying $A \times x + 1 - x   \leq e^{\epsilon}$. If $0 < x \leq \frac{1}{1+e^{\epsilon}}$, we obtain from $A \leq g(x) = g_1(x)\leq g_2(x)$ that $A \leq \frac{e^{\epsilon}-1+x}{x}$,  yielding $A \times x + 1 - x   \leq e^{\epsilon}$. If $x=0$, we have $A \times x + 1 - x  =1  \leq e^{\epsilon}$. To summarize, for any $x \in [0,1]$, it follows that $1- x \geq A \big(e^{-\epsilon} - x\big)$.
\end{itemize}

(\ref{equiva3x}) is proved since its ``$\Longrightarrow$'' and ``$\Longleftarrow$'' both hold.
\pfe

\subsubsection{From membership privacy to Bayesian differential privacy}

\begin{thm} \label{lemMPtoBDP}
For a family of distributions that are downward scalable in the sense of Li \emph{et al}. \cite{Li-CCS2013}, $\epsilon$-membership privacy implies $\epsilon$-Bayesian differential privacy.
\end{thm}

\noindent \textbf{Proof of Theorem~\ref{lemMPtoBDP}.}  The proof is similar to that of \cite[Theorem 3.6]{Li-CCS2013}. For completeness, we still present the details below.

Assume, for the sake of contradiction, that mechanism ${Y}$ achieves ${\epsilon}$-membership privacy yet does not satisfy $\epsilon$-Bayesian differential privacy. Then there
exists a distribution $\mathcal{D}$ and entity $x_i$ such that $0 < \mathbb{P}_{\mathcal{D}}[x_i] < 1$
and $\mathbb{P}_{\mathcal{D},Y}[\mathcal{Y}\con x_i]>e^{\epsilon}\mathbb{P}_{\mathcal{D},Y}[\mathcal{Y}\con \overline{x_i}]$. We discuss two cases below.

Case one: $ \mathbb{P}_{\mathcal{D},Y}[\mathcal{Y}\con \overline{x_i}]= 0$ and $\mathbb{P}_{\mathcal{D},Y}[\mathcal{Y}\con x_i] > 0$. Since $\mathbb{D}$ is downward scalable, by definition $\mathbb{D}$ contains some $\mathbb{D}'$ which is $x_i$-scaled
from $\mathbb{D}$ such that $\mathbb{P}_{\mathcal{D}'}[x_i]<e^{-\epsilon}$. From \cite[Lemma 3.4]{Li-CCS2013}, we have $\mathbb{P}_{\mathcal{D}',Y}[\mathcal{Y}\con \overline{x_i}] = \mathbb{P}_{\mathcal{D},Y}[\mathcal{Y}\con \overline{x_i}] $, which with the case condition $  \mathbb{P}_{\mathcal{D},Y}[\mathcal{Y}\con \overline{x_i}] = 0$ means $\mathbb{P}_{\mathcal{D}',Y}[\mathcal{Y}\con \overline{x_i}]  = 0$, further yielding $\mathbb{P}_{\mathcal{D}',Y}[x_i \con \mathcal{Y}] = 1$. Therefore, $\mathbb{P}_{\mathcal{D}',Y}[x_i \con \mathcal{Y}] = 1>e^{\epsilon}\mathbb{P}_{\mathcal{D}'}[x_i]$, which contradicts the fact that ${Y}$ achieves ${\epsilon}$-membership privacy.

Case two: $\mathbb{P}_{\mathcal{D},Y}[\mathcal{Y}\con x_i]=\alpha\mathbb{P}_{\mathcal{D},Y}[\mathcal{Y}\con \overline{x_i}]$, where $\alpha > e^{\epsilon}$.
Since $\mathbb{D}$ is downward scalable, by definition $\mathbb{D}$ contains some $\mathbb{D}'$ which is $x_i$-scaled
from $\mathbb{D}$ such that $\mathbb{P}_{\mathcal{D}'}[x_i]=q$ for an arbitrarily small
$q$ (see \cite{Li-CCS2013} for the meaning of ``*-scaled''). From \cite[Lemma 3.4]{Li-CCS2013}, we have $\mathbb{P}_{\mathcal{D}',Y}[\mathcal{Y}\con {x_i}] = \mathbb{P}_{\mathcal{D},Y}[\mathcal{Y}\con {x_i}] $ and $\mathbb{P}_{\mathcal{D}',Y}[\mathcal{Y}\con \overline{x_i}] = \mathbb{P}_{\mathcal{D},Y}[\mathcal{Y}\con \overline{x_i}] $. These with the case condition $\mathbb{P}_{\mathcal{D},Y}[\mathcal{Y}\con x_i]=\alpha\mathbb{P}_{\mathcal{D},Y}[\mathcal{Y}\con \overline{x_i}]$ gives $\mathbb{P}_{\mathcal{D}',Y}[\mathcal{Y}\con x_i]=\alpha\mathbb{P}_{\mathcal{D}',Y}[\mathcal{Y}\con \overline{x_i}]$. Then,  under $\mathcal{D}'$, we
have
\begin{align}
\frac{\mathbb{P}_{\mathcal{D}',Y}[x_i \con \mathcal{Y}]}{\mathbb{P}_{\mathcal{D}'}[x_i]}   &=\frac{\mathbb{P}_{\mathcal{D}',Y}[\mathcal{Y} \con x_i]}{\mathbb{P}_{\mathcal{D}',Y}[\mathcal{Y}]} \nonumber \\ &=\frac{\mathbb{P}_{\mathcal{D}',Y}[\mathcal{Y} \con x_i]}{\mathbb{P}_{\mathcal{D}',Y}[\mathcal{Y} \con x_i]\mathbb{P}_{\mathcal{D}'}[x_i]+\mathbb{P}_{\mathcal{D}',Y}[\mathcal{Y} \con \overline{x_i}]\mathbb{P}_{\mathcal{D}'}[\overline{x_i}]} \nonumber \\ & =\frac{\alpha\mathbb{P}_{\mathcal{D}',Y}[\mathcal{Y}\con \overline{x_i}]}{\alpha\mathbb{P}_{\mathcal{D}',Y}[\mathcal{Y}\con \overline{x_i}]\cdot q+\mathbb{P}_{\mathcal{D}',Y}[\mathcal{Y} \con \overline{x_i}]\cdot (1-q)}  \nonumber \\ &= \frac{\alpha}{\alpha q + 1-q}.
\end{align}
The above ratio $\frac{\alpha}{\alpha q + 1-q}$ is greater than $e^{\epsilon}$ given $\alpha > e^{\epsilon}$, once we ensure $q<\frac{\alpha - e^{\epsilon}}{e^{\epsilon}(\alpha -1)}$. This will give
$\mathbb{P}_{\mathcal{D}',Y}[x_i \con \mathcal{Y}] >e^{\epsilon}\mathbb{P}_{\mathcal{D}'}[x_i]$, which contradicts the fact that ${Y}$ achieves ${\epsilon}$-membership privacy.

Summarizing the above two cases, we have proved the desired result. \pfe

\section{Related Work}  \label{related}

 The notion of \emph{differential privacy} (DP)  \cite{Dwork2006,dwork2006calibrating} provides a rigorous foundation for privacy protection. Intuitively, DP implies that changing one entry in the
database does not significantly change the query output, so that an adversary, seeing the query output and knowing all records except the one to be inferred,  draws almost the same conclusion on whether or not a record is in the database. Differential privacy has received considerable interest in the literature \cite{wang2017privsuper,6686180,zhang2015private,tramer2015differential,qin2016heavy,Jiang:2013:PTD:2484838.2484846,erlingsson2014rappor,acs2017differentially,zhao2017preserving,mohasselsecureml}. Yang \textit{et al.}  \cite{yang2015bayesian} and  Liu~\textit{et~al.}~\cite{Changchang2016} propose  Bayesian differential privacy  and dependent differential privacy respectively to generalize differential privacy for correlated data.
 Kasiviswanathan and Smith~\cite{kasiviswanathan2014semantics}
 propose a Bayesian formulation of semantic privacy, inspired by the following interpretation of differential privacy explained in~\cite{Dwork2006}: \emph{Regardless of external knowledge, an adversary
with access to the sanitized database draws the same conclusions whether or not any individual data is included in the original database}. To present the notion of semantic privacy, Kasiviswanathan and Smith model the external knowledge
 via a prior probability distribution, and model  conclusions  via the corresponding posterior distribution.  Li~\textit{et~al.}~\cite{Li-CCS2013} introduce membership privacy (MP) in  consideration of the adversary's prior beliefs.

Dwork and Rothblum \cite{dwork2016concentrated} recently proposed the notion of \textit{concentrated differential privacy}, a relaxation of differential privacy achieving
better accuracy than     differential privacy without
compromising on cumulative privacy cost over multiple computations. Motivated by \cite{dwork2016concentrated}, Bun and Steinke \cite{bun2016concentrated} suggest a relaxation of
concentrated differential privacy. Instead of treating the privacy loss as a subgaussian random
variable as \cite{dwork2016concentrated} does, Bun and Steinke \cite{bun2016concentrated} instead formulate the problem in terms of Renyi entropy, giving a
relaxation of concentrated differential privacy. Jorgensen \textit{et al.} \cite{jorgensen2015conservative} introduce  a new privacy definition called personalized differential
privacy, a generalization of differential privacy in which
users specify a personal privacy level for their data. They show that by accepting that not all users demand the same
level of privacy, a higher level of utility can often be obtained by
not providing excess privacy budget to those who do not need it. They present a mechanism for achieving
personalized differential
privacy, inspired by the well-known exponential mechanism of differential
privacy. Hall \textit{et al.} \cite{hall2012random} introduce additional randomness to extend differential
privacy to the notion of random differential
privacy. Compared with differential
privacy,  Lee and Clifton \cite{lee2012differential} give an alternate formulation,
differential identifiability, parameterized by the
probability of individual identification. Their notion provides the
strong privacy guarantees of differential privacy, while allowing
policy makers to set parameters based on the
privacy concept of individual identifiability.

Bohli and Andreas \cite{bohli2011relations} discuss the relations among several privacy definitions, but the discussion does not cover differential privacy. Li \textit{et al.} \cite{li2012sampling} present the relation between $k$-anonymization and differential privacy, where the $k$-anonymity notion by \cite{sweeney2002k,samarati2001protecting} means that when only quasi-identifiers are considered, each
record in a
$k$-anonymized dataset should appear at least
$k$ times.
 Wang \textit{et al.} \cite{wang2016relation} analyze the relation between differential privacy, mutual-information privacy, and identifiability.
 Mironov \mbox{\textit{et al.}} \cite{mironov2009computational} present several relaxations of differential privacy by requiring
privacy guarantees to hold only against computationally bounded adversaries. They establish various relations among these
notions, and show that the notions exhibit close connection with the theory
of pseudodense sets  \cite{reingold2008dense}.

\section{Conclusion}
\label{sec:Conclusion}

In this paper, we present a comprehensive view of the relations among different privacy notions: differential privacy (DP), Bayesian differential privacy (BDP), semantic privacy (SP), and membership privacy (MP). In particular, we extend the notion of semantic privacy (SP) to Bayesian semantic privacy (BSP) and prove its essential equivalence with Bayesian differential privacy (BDP) in the quantitative sense. We show the relations between BDP, BSP, and SP as follows: \mbox{$\epsilon$-BDP $\Longleftarrow$ $\big(\frac{1}{2}-\frac{1}{e^{\epsilon}+1}\big)$-BSP}, and \mbox{$\epsilon$-BDP $\Longrightarrow$ $(e^{2\epsilon}-1)$-BSP $\Longrightarrow$ $(e^{2\epsilon}-1)$-SP}. Moreover, we \mbox{derive} the following  relations between BDP and MP.
First, \mbox{$\epsilon$-BDP $\Longrightarrow$ $\epsilon$-MP}.
Second, For a family of distributions that are downward scalable in the sense of \mbox{Li \emph{et al}. \cite{Li-CCS2013}, it holds that $\epsilon$-BDP $\Longleftarrow$ $\epsilon$-MP.}

\scriptsize

\let\OLDthebibliography\thebibliography
 \renewcommand\thebibliography[1]{
   \OLDthebibliography{#1}
   \setlength{\parskip}{.5pt}
   \setlength{\itemsep}{1pt plus 0ex}
 }


\normalsize

\end{document}